\documentclass[11pt]{article}
\usepackage{amssymb,amsmath,amsthm,amsfonts}
\usepackage{jinshuomacros}
\usepackage{algorithm}
\usepackage[noend]{algpseudocode}
\usepackage{thmtools}
\usepackage{mathrsfs}
\usepackage[usenames,dvipsnames]{color}
\usepackage{tikz}
\usepackage[hidelinks]{hyperref}
\usepackage{natbib}

\usepackage{graphicx}
\usepackage{bibentry}
\usepackage{cleveref}
\usepackage{fullpage}

\newif\ifhideproofs

\ifhideproofs
\usepackage{environ}
\NewEnviron{hide}{}

\fi

\nobibliography* 

\declaretheoremstyle[bodyfont=\normalfont]{normalbody}

\begin{document}
\title{Rejoinder: Gaussian Differential Privacy}


\author{Jinshuo Dong\thanks{Graduate Group in Applied Mathematics and Computational Science, University of Pennsylvania. Email: \texttt{jinshuo@sas.upenn.edu}.} \and Aaron Roth\thanks{Department of Computer and Information Sciences, University of Pennsylvania. Email: \texttt{aaroth@cis.upenn.edu}.} \and Weijie J.~Su\thanks{Department of Statistics, the Wharton School, University of Pennsylvania. Email: \texttt{suw@wharton.upenn.edu}.}}



\date{April 2021}


\maketitle

We warmly thank Editor Paul Smith for selecting our paper for discussion and are extremely grateful to all the discussants for taking their valuable time to provide engaging and stimulating feedback on our work. These insights situate our work in context and provide promising directions for future research. We are excited to see that thoughts about theoretical complements and new applications are already  emerging.

A general view, shared by all discussants, is that privacy is a first-order concern in many data science problems. We are very pleased to learn that our statistics community  welcomes new foundational development and methodological contributions that allow for privacy protections in statistical data analysis.

In this rejoinder, we aim to address two broad issues that cover most comments made in the discussion. First, we discuss some theoretical aspects of our work and comment on how this work might impact the theoretical foundation of privacy-preserving data analysis. Taking a practical viewpoint, we next discuss how $f$-differential privacy ($f$-DP) and Gaussian differential privacy (GDP) can make a difference in a range of applications.


\section{Theoretical Aspects of $f$-DP}
\label{sec:theor-aspects-f}


As echoed by many discussants, the formalization of $f$-DP in our paper \citep{dong2019gaussian} starts from a decision-theoretic interpretation of a ``differential'' privacy attack, which originates from the work of \citet{wasserman2010statistical}. The binary nature of the decision-theoretic problems renders the classical theory of hypothesis testing a basic tool. Specifically, we use the trade-off between type I and type II errors of the hypothesis testing problem as our privacy measure. In response to a question raised by Dr.~Jorge Mateu, we remark that this trade-off is concerned with a thought experiment in which someone is trying to determine if an individual's data is in the dataset or not, rather than a hypothesis test that is actually conducted. It can therefore always be reasoned about analytically/formally, without needing simulation, even if the algorithms themselves are complex or simulation based.


This treatment of  privacy cost in  $f$-DP comes with several technical properties that can facilitate the development of better differentially private algorithms. As highlighted by Dr.~Borja Balle, $f$-DP gives tight and analytically tractable formulas for composition. This appealing feature arises from applying the central limit theorem to the privacy loss random variables, thereby making GDP a canonical single-parameter family of privacy definitions within the $f$-DP class. While we did not attempt to push hard on weakening assumptions for the privacy central limit theorems, there are several possible extensions. For example, one may be able to identify a necessary and sufficient condition for the privacy central limit theorem to hold, just like the Lindeberg--Feller condition for the usual central limit theorems. Another possibility is to sharpen the central limit theorem by
leveraging a refined analysis of the privacy loss random variables~\citep{zheng2020sharp}. More specifically, Dr.~Sebastian Dietz suggested a very interesting direction for improving the composition formulas by making use of a third-moment-free central limit theorem
(see, for example, \cite{dasgupta2008asymptotic}). A successful investigation in this direction might extend the applicability of the composition formulas to $(\epsilon, \delta)$-DP and others. More broadly, it would also be interesting to explore central limit theorem phenomena of privacy beyond composition. For example, \cite{dong2021} recently showed that a related central limit theorem occurs in high-dimensional query answering and yet privacy cost is best described in the framework of $f$-DP. We see all these as interesting future directions for broadening the scope of the hypothesis testing viewpoint on differential privacy.


In addition to composition, subsampling is  another important primitive that is involved in many algorithm designs. As pointed out by Dr.~Borja Balle, while divergence-based privacy definitions face technical difficulties in describing privacy amplification by subsampling, $f$-DP gives a relatively concise and coherent expression for understanding how privacy is amplified using this primitive. This also gives a sharper privacy analysis of subsampling than can be obtained by directly using $(\epsilon, \delta)$-DP. An interesting observation made by Dr.~Borja Balle is that the significant gap between the two frameworks seems surprising, and warrants further investigation. It is also worth developing similar privacy analyses for the various flavors of  subsampling schemes used in training deep learning models (not all of which involve \emph{independent} sampling across rounds).


\section{Applications of $f$-DP}
\label{sec:applicability-f-dp}


Our main hope for $f$-DP is to see as many applications as possible  to improving the privacy analyses of the diverse algorithms used in a variety of data science problems. Encouragingly, we found many such possibilities in the discussants' contributions that either tackle important problems or show great promise.

Understanding the tradeoff between privacy and utility for various statistical and computational tasks is \emph{the} central object of study in the differentially privacy literature. The main point of $f$-DP and GDP is to make it possible to capture this fundamental tradeoff more precisely. As a result, the $f$-DP framework allows us to obtain better tradeoffs between privacy guarantees and data usability. This tradeoff is different in different applications, and requires analyses on a case by case basis. We emphasize, as pointed out by several of the discussants, that  privacy protection does inevitably come with utility loss. Indeed, this is a consequence of the ``fundamental law of information recovery'', which applies not just to differential privacy but to any method of
releasing data. So while it is true that differential privacy can harm utility (especially for small datasets), this is not an artifact of differential privacy, but an actual, fundamental tradeoff that we have to grapple with as a society. We can choose to get exact statistics about our data, but we should understand that this means giving up on privacy. Differential privacy takes no stand on how we should mediate this fundamental tradeoff: rather it provides a precise language in which to talk about it.


\subsection{$f$-DP for Stochastic Optimization}
\label{sec:f-dp-machine}

To appreciate how sharply this tradeoff can be characterized using a given privacy definition, perhaps the best benchmark is stochastic gradient descent (SGD), the basic foundation for many machine learning algorithms. Owing to its effectiveness in handling composition and subsampling, $f$-DP gives a tighter privacy analysis of SGD than the moments accountant technique~\citep{abadi2016deep}, which further feeds back into improved test accuracy of trained deep learning models at fixed privacy guarantees~\citep{bu2020deep}. We are delighted that Dr.~Borja Balle wrote ``I would encourage practitioners to take note of this and start using Gaussian DP accounting in their DP-SGD implementations.''

Moving forward, Dr.~Marco Avella-Medina raised several interesting and important questions regarding private SGD with $f$-DP guarantees. Although gradient clipping is a necessary step in private SGD that ensures bounded sensitivity to any single data point, this step can lead to inconsistency for some problems. To go around this difficulty, Avella-Medina suggested using a consistent bounded influence $M$-estimator from robust statistics, which we believe is a promising approach worthy of future research effort. Moreover, we are glad to see that \citet{avella2021differentially} introduced a kind of noisy gradient descent and analyzed its Gaussian differential privacy properties. This opens an exciting research avenue to understand when noisy gradient descent outperforms SGD.


\subsection{Other Applications}
\label{sec:f-dp-medical}

Differential privacy has applications beyond machine learning. A promising application area --- due to strict privacy regulation --- is in the analysis and sharing of medical data. A challenge in  medical  applications is that the size of the relevant datasets is often relatively small. The improved tradeoff between privacy and utility is especially important in the challenging small data regime. As noted by Drs.~Peter Krusche and Frank Bretz, there are obstacles to combining data across hospitals --- for which we think differential privacy might be able to help. Noisy access to a large dataset might be better --- even from the perspective of utility --- than exact access to only a small local dataset. When applying privacy protections to small data, it is especially important not be as tight as possible in accounting for privacy loss, which is one of the main benefits of the $f$-DP framework.

Dr.~Jorge Mateu brought up  privacy issues that arise when analyzing trajectory data. In principle, $f$-DP and GDP can be applied to any kind of data, such as trajectory data. It would make sense, for example, to think about the question of releasing statistics about trajectory data or a synthetic dataset consisting of trajectories that maintain consistency with the real data with respect to various statistics of interest, so long as those statistics have low sensitivity and vary only mildly with the data of individuals. These types of problems deserve further study. Of course, providing useful analyses of a single \emph{individual}'s trajectory is by design prevented by technologies that aim to preserve individual privacy. A related question, asked by Dr.~Christine Chai, was whether the $f$-DP framework can be applied to COVID-19 contact tracing data. Differential privacy ($(\epsilon, \delta)$-DP, $f$-DP, or any related variant) is not directly applicable to what is most commonly known as contact tracing --- letting contacts know that someone with COVID-19 has been near them --- since by design, this is highly sensitive to a single
data point. However, GDP (as well as other differential privacy variants) can be used to improve population level statistics related to contact tracing, such as how crowded grocery stores are by time and mobility data, or even what fraction of visitors to a grocery store in a given day have had potential COVID-19 exposure. More generally, we believe that $f$-DP has many more connections to various aspects of data science.


Finally, we remark that there are many heuristic approaches to privacy that do not come with the guarantees of differential privacy. There is a vast literature of pros and cons among these approaches, which is beyond the scope of this paper --- but in general, ``syntactic'' approaches do not stand up to attack by a determined adversary. In particular, synthetic data is known to be neither necessary nor sufficient for privacy --- but also not incompatible with differential privacy. For example, there is a large literature on generating differentially private synthetic data (see, e.g., \cite{syn1,syn2,syn3,syn4,syn5,syn6}), most of which we believe can be improved by $f$-DP style analyses.

\bibliographystyle{abbrvnat}
\bibliography{ref}

\end{document}